\documentclass[sigconf]{acmart}

\usepackage{url}
\usepackage{balance}
\usepackage{enumitem}
\usepackage{caption}	
\usepackage{multirow}
\usepackage{multicol}
\usepackage{amsmath}
\usepackage{verbatim}
\usepackage{float}
\usepackage[normalem]{ulem}
\usepackage{xspace}
\usepackage{listings}
\usepackage{color}
\usepackage{subcaption}

\newcommand{\ie}{\textit{i.e.,}\xspace}
\newcommand{\eg}{\textit{e.g.,}\xspace}
\newcommand{\etc}{\textit{etc.}\xspace}
\newcommand{\etal}{\textit{et al.}\xspace}
\newcommand{\secref}[1]{Section~\ref{#1}\xspace}

\newcommand{\figref}[1]{Figure~\ref{#1}\xspace}

\newcommand{\tabref}[1]{Table~\ref{#1}\xspace}

\newcommand{\mplus}{{\sc MDroid+}\xspace}
\newcommand{\alltoolsopers}{102\xspace}
\newcommand{\numDocs}{2,023\xspace} 
\newcommand{\numFaults}{262\xspace} 
\newcommand{\numCategs}{14\xspace} 
\newcommand{\numMuts}{38\xspace} 
\newcommand{\numImplMuts}{35\xspace} 

\include{macro}

\setcopyright{acmcopyright}
\acmDOI{10.1145/3106237.3106275}
\acmISBN{978-1-4503-5105-8/17/09}
\acmConference[ESEC/FSE'17]{2017 11th Joint Meeting of the European Software Engineering Conference and the ACM SIGSOFT Symposium on the Foundations of Software Engineering}{September 4--8, 2017}{Paderborn, Germany} 
\acmYear{2017}
\copyrightyear{2017}
\acmPrice{15.00}

\begin{document}

\title{Enabling Mutation Testing for Android Apps}
\author{Mario Linares-V\'{a}squez}
\affiliation{Universidad de los Andes \\ Colombia}
\author{Gabriele Bavota}
\affiliation{Universit\`{a} della Svizzera italiana \\ Switzerland}
\author{Michele Tufano}
\affiliation{College of William and Mary \\United States}
\author{Kevin Moran}
\affiliation{College of William and Mary \\United States}
\author{Massimiliano Di Penta}
\affiliation{University of Sannio \\ Italy}
\author{Christopher Vendome}
\affiliation{College of William and Mary \\United States}
\author{Carlos Bernal-C\'ardenas}
\affiliation{College of William and Mary \\United States}
\author{Denys Poshyvanyk}
\affiliation{College of William and Mary \\United States}


  \renewcommand{\shortauthors}{M.~Linares-V\'{a}squez, G.~Bavota, M.~Tufano, K.~Moran,\\
	M.~Di~Penta, C.~Vendome, C.~Bernal-C\'ardenas, and D.~Poshyvanyk}

\begin{abstract}
Mutation testing has been widely used to assess the fault-detection effectiveness of a test suite, as well as to guide test case generation or prioritization. Empirical studies have shown that, while mutants are generally representative of real faults, an effective application of mutation testing requires ``traditional" operators designed for programming languages to be augmented with operators specific to an application domain and/or technology.
This paper proposes \mplus, a framework for effective mutation testing of Android apps. First, we systematically devise a taxonomy of \numFaults types of Android faults grouped in \numCategs categories by manually analyzing \numDocs software artifacts from different sources (\eg bug reports, commits). Then, we identified a set of \numMuts mutation operators, and implemented an infrastructure to automatically seed mutations in Android apps with 35 of the identified operators. The taxonomy and the proposed operators have been evaluated in terms of stillborn/trivial mutants generated and their capacity to represent real faults in Android apps, as compared to other well know mutation tools.

\end{abstract}

\maketitle

\section{Introduction}
\label{sec:intro}
In the last few years mobile apps have become indispensable in our daily lives. With millions of mobile apps available for download on Google Play~\cite{GP} and the Apple App Store~\cite{AppleStore}, mobile users have access to an unprecedentedly large set of apps that are not only intended to provide entertainment but also to support critical activities such as banking and health monitoring. Therefore, given the increasing relevance and demand for high quality apps, industrial practitioners and academic researchers have been devoting significant effort to improving methods for measuring and assuring the quality of mobile apps. Manifestations of interest in this topic include the broad portfolio of mobile testing methods ranging from tools for assisting record and replay testing~\cite{Gomez:ICSE13,Hu:OOPSLA2015}, to automated approaches that generate and execute test cases~\cite{Machiry:FSE13,Shauvik:2015,Linares:MSR15,Moran:ICST16,Mao:ISSTA16}, and cloud-based services for large-scale multi-device testing \cite{Perfecto}.

Despite the availability of these tools/approaches, the field  of mobile app testing is still very much under development; as evidenced by limitations related to test data generation \cite{Shauvik:2015,Linares-Vasquez:ICSME17}, and concerns regarding effective assessment of the quality of mobile apps' test suites. One way to evaluate test suites is to seed small faults, called mutants, into source code and asses the ability of a suite to detect these faults \cite{Hamlet:TSE,DeMillo:Computer}. Such mutants have been defined in the literature to reflect the typical errors developers make when writing source code \cite{Ma:ISSRE03,KimCM0,Moeller93,Ostrand:2002:DFL:566171.566181,Zhang08,Nistor:2013:DRF:2487085.2487134, Robinson2009}. 

However, existing literature lacks a thorough characterization of bugs exhibited by mobile apps. Therefore, it is unclear whether such apps exhibit a distribution of faults similar to other systems, or if there are types of faults that require special attention. As a consequence, it is unclear whether the use of traditional mutant taxonomies \cite{Ma:ISSRE03,KimCM0} is enough to asses test quality and drive test case generation/selection of mobile apps.

In this paper, we explore this topic focusing on apps developed for Android, the most popular mobile operating system. Android apps are characterized by GUI-centric design/interaction, event-driven programming, Inter Processes Communication (IPC), and interaction with backend and local services. In addition, there are specific characteristics of Android apps---such as permission mechanisms, Software Development Kit (SDK) version compatibility, or features of target devices---that can lead to a failure. While this set of characteristics would demand a specialized set of mutation operators that can support mutation analysis and testing, there is no available tool to date that supports mutation analysis/testing of Android apps, and relatively few (eight) mutation operators have been proposed by the research community~\cite{Deng:ICSTW15}. At the same time, 
 mutation tools for Java apps, such as Pit \cite{PIT} and Major \cite{Just:ISSTA14,Just:SK2011} lack any Android-specific mutation operators, and present challenges for their use in this context, resulting in common problems such as trivial mutants that always crash at runtime or difficulties automating mutant compilation into Android PacKages (APKs).
 

{\bf Paper contributions.} This paper aims to deal with the lack of (i) an extensive empirical evidence of the distribution of Android faults, (ii) a thorough catalog of Android-specific mutants, and (iii) an analysis of the applicability of state-of-the-art mutation tools on Android apps. We then propose a framework, \mplus, that relies on a catalog of mutation operators inspired by a taxonomy of bugs/crashes specific for Android apps, and a profile of potential failure points automatically extracted from APKs. 
 
As a first step, we produced a taxonomy of Android faults by analyzing a statistically significant sample of  \numDocs candidate faults documented in (i) bug reports from open source apps, (ii)  bug-fixing commits of open source apps;  (iii) Stack Overflow discussions, (iv) the Android exception hierarchy and APIs potentially triggering such exceptions; and (v) crashes/bugs described in previous studies on Android~\cite{Machiry:FSE13,Ravindranath:MobiSys14,Zhang:TOSEM14,Zaeem:ICST2014,Liang:MobiCom14,Adamsen:ISSTA15,Shauvik:2015,Moran:FSE15,Moran:ICST16}.  As a result, we produced a taxonomy of \numFaults types of faults grouped in \numCategs categories, four of which relate to Android-specific faults, five to Java-related faults, and five mixed categories (\figref{fig:taxonomy}). Then, leveraging this fault taxonomy and focusing on Android-specific faults, we devised a set of \numMuts Android mutation operators and implemented a  platform to automatically seed \numImplMuts of them. Finally, we conducted a study comparing \mplus with other Java and Android-specific mutation tools. The study results indicate that \mplus, as compared to existing competitive tools, (i) is able to cover a larger number of bug types/instances present in Android app, (ii) is highly complementary to the existing tools in terms of covered bug types, and (iii) generates fewer trivial and stillborn mutants.

\section{Related Work}
\label{sec:related}
This section describes related literature and publicly available, state-of-the-art tools on mutation testing. We do not discuss the literature on testing Android apps \cite{Amalfitano:ASE12,Hao:MobiSys14,Mahmood:FSE14,Machiry:FSE13,Shauvik:2015,Linares:MSR15,Moran:ICST16,Mao:ISSTA16,Linares-Vasquez:ICSME17}, since proposing a novel approach for testing Android apps is not the main goal of this work.
For further details about the concepts, recent research, and future work in the field of mutation testing, one can refer to the survey by Jia and Harman~\cite{Jia:TSE}.

\textbf{Mutation Operators.} Since the introduction of mutation testing in the 70s~\cite{Hamlet:TSE,DeMillo:Computer}, researchers have tried not only to define new mutation operators for different programming languages and paradigms (\eg mutation operators have been defined for Java \cite{Ma:ISSRE03} and Python \cite{Derezinska2014}) but also for specific types of software like Web applications \cite{Praphamontripong:Mutation15} and data-intensive applications \cite{Appelt:ISSTA14,ZhouF09} either to exercise their GUIs \cite{Oliveira:ICSTW15} or to alter complex, model-defined input data \cite{Nardo:ICST15}.
The aim of our research, which we share with prior work, is to define customized mutation operators suitable for Android applications, by relying on a solid empirical foundation. 

To the best of our knowledge, the closest work to ours is that of Deng \etal, \cite{Deng:ICSTW15}, which defined eight mutant operators aimed at introducing faults in the essential programming elements of Android apps, \ie intents, event handlers, activity lifecycle, and XML files (\eg GUI or permission files). While we share with Deng \etal the need for defining specific operators for the key Android programming elements, our work builds upon it by (i) empirically analyzing the distribution of faults in Android apps by manually tagging \numDocs documents, (ii) based on this distribution, defining a mutant taxonomy---complementing Java mutants---which includes a total of \numMuts operators tailored for the Android platform. 

\textbf{Mutation Testing Effectiveness and Efficiency.}
Several researchers have proposed approaches  to measure the effectiveness and efficiency of mutation testing~\cite{Offutt:1991,Andrews:ICSE05,Gopinath:ICSTW16,Just:ICST12}  to devise strategies for reducing the effort required to generate effective mutant sets~\cite{Adamopoulos:2004,Just:ISSTA14-2,Gopinath:ICSE16}, and to define theoretical frameworks~\cite{Jia:TSE,Shin:ICST16}. Such strategies can complement our work, since in this paper we aim at defining new mutant operators for Android, on which effectiveness/efficiency measures or minimization strategies can be applied.

\textbf{Mutation Testing Tools.} Most of the available mutation testing tools are in the form of research prototypes. Concerning Java, representative tools are $\mu$Java~\cite{Ma:2005},  Jester~\cite{Jester},  Major~\cite{Just:ISSTA14}, Jumble~\cite{Jumble}, PIT~\cite{PIT}, and javaLanche~\cite{Schuler:FSE09}. Some of these tools operate on the Java source code, while others inject mutants in the bytecode. For instance, $\mu$Java,  Jester, and Major generate the mutants by modifying the source code, while Jumble, PIT, and javaLanche perform the mutations in the bytecode. When it comes to Android apps, there is only one available tool, namely muDroid \cite{muDroid}, which performs the mutations at byte code level by generating one APK (\ie one version of the mobile app) for each mutant. The tools for mutation testing can be also categorized according to the tool's capabilities (\eg the availability of automatic tests selection). A thorough comparison of these tools is out of the scope of this paper. The interested reader can find more details on PIT's website \cite{MutationTools} and in the paper by Madeysky and Radyk~\cite{Madeyski:IET10}.

\textbf{Empirical Studies on Mutation Testing.} 
Daran and Th\'evenod-Fosse \cite{DaranT96} were the first to empirically compare mutants and real faults, finding that the set of errors and failures they produced with a given test suite were similar. Andrews \etal \cite{Andrews:ICSE05,AndrewsBLN06} studied whether mutant-generated faults and faults seeded by humans can be representative of real faults. The study showed that carefully-selected mutants are not easier to detect than real faults, and can provide a good indication of test suite adequacy, whereas human-seeded faults can likely produce underestimates. Just \etal \cite{Just:FSE14} correlated mutant detection and real fault detection using automatically and manually generated test suites. They found that these two variables exhibit a statistically significant correlation. At the same time, their study pointed out that traditional Java mutants need to be complemented by further operators, as they found that around 17\% of faults were not related to mutants.
 

\section{A Taxonomy of Crashes/Bugs in Android apps}
\label{sec:taxonomy}
To the best of our knowledge there is currently no (i) large-scale study describing a taxonomy of bugs in Android apps, or (ii) comprehensive mutation framework including operators derived from such a taxonomy and targeting mobile-specific faults (the only framework available is the one with eight mutation operators proposed by Deng \etal \cite{Deng:ICSTW15}).  In this section, we describe a taxonomy of bugs in Android apps derived from a large manual analysis of (un)structured sources. Our work is the first large-scale data driven effort to design such a taxonomy. Our purpose is to extend/complement previous studies analyzing bugs/crashes in Android apps and to provide a large taxonomy of bugs that can be used to design mutation operators. In all the cases reported below the manually analyzed sets of sources---randomly extracted---represent a 95\% statistically significant sample with a 5\% confidence interval.

\subsection{Design}
\label{sub:taxdesing}
To derive such a taxonomy we manually analyzed six different sources of information described below:
\begin{enumerate}[leftmargin=*]
\item \emph{Bug reports of Android open source apps}. Bug reports are the most obvious source to mine in order to identify typical bugs affecting Android apps. We mined the issue trackers of 16,331 open source Android apps hosted on GitHub. Such apps have been identified by locally cloning all Java projects (381,161) identified through GitHub's API and searching for projects with an \textit{AndroidManifest.xml} file (a requirement for Android apps) in the top-level directory. We then removed forked projects to avoid duplicated apps and filtered projects that did not have a single star or watcher to avoid abandoned apps. We utilized a web crawler to mine the GitHub issue trackers. To be able to analyze the bug cause, we only selected closed issues (\ie those having a fix that can be inspected) having ``Bug'' as type. Overall, we collected 2,234 issues from which we randomly sampled 328 for manual inspection. 

\item \emph{Bug-fixing commits of Android open source apps}. Android apps are often developed by very small teams \cite{Mona:ESEM13,Nagappan:SANER16}. Thus, it is possible that some bugs are not documented in issue trackers but quickly discussed by the developers and then directly fixed. This might be particularly true for bugs having a straightforward solution. Thus, we also mined the versioning system of the same 16,331 Android apps considered for the bug reports by looking for bug-fixing commits not related to any of the bugs considered in the previous point (\ie the ones documented in the issue tracker). With the cloned repositories, we utilized the \emph{git} command line utility to extract the commit notes and matched the ones containing lexical patterns indicating bug fixing activities, \eg {\em ``fix issue''}, {\em ``fixed bug''}, similarly to the approach proposed by Fischer \etal \cite{FischerPG03}. By exploiting this procedure we collected 26,826 commits, from which we randomly selected a statistically significant sample of 376 commits for manual inspection.

\item \emph{Android-related Stack Overflow (SO) discussions}. It is not unusual for developers to ask help on SO for bugs they are experiencing and having difficulty fixing~\cite{Linares:MSR13,Beyer:ICSME14,Khalid:IEEE2014,Rosen:EMSE16}. Thus, mining SO discussions could provide additional hints on the types of bugs experienced by Android developers. To this aim, we collected all 51,829 discussions tagged ``Android'' from SO. Then, we randomly extracted a statistically significant sample of  377 of them for the manual analysis.

\item \emph{The exception hierarchy of the Android APIs}. Uncaught exceptions and statements throwing exceptions are a major source of faults in Android apps~\cite{Zhang:TOSEM14,Coelho:MSR15}. We automatically crawled the official Android developer JavaDoc guide to extract the exception hierarchy and API methods throwing exceptions. We collected 5,414 items from which we sampled 360 of them for manual analysis.

\item \emph{Crashes/bugs described in previous studies on Android apps}. 43 papers related to Android testing\footnote{The complete list of papers is provided with our online appendix \cite{replication}.} were analyzed by looking for crashes/bugs reported in the papers. For each identified bug, we kept track of the following information: app, version, bug id, bug description, bug URL. When we were not able to identify some of this information, we contacted the paper's authors. In the 43 papers, a total of 365 bugs were mentioned/reported; however, we were able (in some cases with the authors' help) to identify the app and the bug descriptions for only 182 bugs/issues (from nine papers \cite{Machiry:FSE13,Ravindranath:MobiSys14,Zhang:TOSEM14,Zaeem:ICST2014,Liang:MobiCom14,Adamsen:ISSTA15,Shauvik:2015,Moran:FSE15,Moran:ICST16}). Given the limited number, in this case we considered all of them in our manual analysis. 

\item \emph{Reviews posted by users of Android apps on the Google Play store}. App store reviews have been identified as a prominent source of bugs and crashes in mobile apps~\cite{Pagano:IREC13,Iacob:MSR13,Khalid:IEEE2014,Panichella:ICSME15,Palomba:ICSME15,Linares-Vasquez:ICSME15}.  However, only a reduced set of reviews are in fact informative and useful for developers \cite{Chen:icse2014,Palomba:ICSME15}. Therefore, to automatically detect informative reviews reporting bugs and crashes, we leverage CLAP, the tool developed by Villarroel \etal \cite{Villarroel:ICSE16}, to automatically identify the bug-reporting reviews. Such a tool has been shown to have a precision of 88\% in identifying this specific type of review. We ran CLAP on the Android user reviews dataset made available by Chen \etal \cite{Chen:2015}. This dataset reports user reviews for multiple releases of $\sim$21K apps, in which CLAP identified 718,132 reviews as bug-reporting. Our statistically significant sample included 384 reviews that we analyzed.
\end{enumerate}

The data collected from the six sources listed above was manually analyzed by the eight authors following a procedure inspired by open coding~\cite{Miles2013}. In particular, the 2,007 documents (\eg bug reports, user reviews, \etc) to manually validate were equally and randomly distributed among the authors making sure that each document was classified by two authors. The goal of the process was to identify the exact reason behind the bug and to define a tag (\eg \emph{null GPS position}) describing such a reason. Thus, when inspecting a bug report, we did not limit our analysis to the reading of the bug description, but we analyzed (i) the whole discussion performed by the developers, (ii) the commit message related to the bug fixing, and (iii) the patch used to fix the bug (\ie source code diff). The tagging process was supported by a Web application that we developed to classify the documents (\ie to describe the reason behind the bug) and to solve conflicts between the authors. Each author independently tagged the documents assigned to him by defining a tag describing the cause behind a bug. Every time the authors had to tag a document, the Web application also shows the list of tags created so far, allowing the tagger to select one of the already defined tags. Although, in principle, this is against the notion of open coding, in a context like the one encountered in this work, where the number of possible tags (\ie cause behind the bug) is extremely high, such a choice helps using consistent naming and does not introduce a substantial bias. 

In cases for which there was no agreement between the two evaluators ($\sim$43\% of the classified documents), the document was automatically assigned to an additional evaluator. The process was iterated until all the documents were classified by the absolute majority of the evaluators with the same tag. When there was no agreement after all eight authors tagged the same document (\eg four of them used the tag $t_{1}$ and the other four the tag $t_{2}$), two of the authors manually analyzed these cases in order to solve the conflict and define the most appropriate tag to assign (this happened for $\sim$22\% of the classified documents). It is important to note that the Web application did not consider documents tagged as \emph{false positive} (\eg a bug report that does not report an actual bug in an Android app) in the count of the documents manually analyzed. This means that, for example, to reach the 328 bug reports to manually analyze and tag, we had to analyze 400 bug reports (since 72 were tagged as false positives).

It is important to point out that, during the tagging, we discovered that for user reviews, except for very few cases, it was impossible (without internal knowledge of an app's source code) to infer the likely cause of the failure (fault) by only relying on what was described in the user review. For this reason, we decided to discard user reviews from our analysis, and this left us with 2,007-384=1,623 documents to manually analyze.

After having manually tagged all the documents (overall, \numDocs = 1,623 + 400 additional documents, since 400 false positives were encountered in the tagging process), all the authors met online to refine the identified tags by merging similar ones and splitting generic ones when needed. Also, in order to build the fault taxonomy, the identified tags were clustered in cohesive groups at two different levels of abstraction, \ie categories and subcategories. Again, the grouping was performed over multiple iterations, in which tags were moved across categories, and categories merged/split.

Finally, the output of this step was (i) a taxonomy of representative bugs for Android apps, and (ii) the assignment of the analyzed documents to a specific tag describing the reason behind the bug reported in the document. 

\begin{figure*}
\includegraphics[width=0.9\linewidth]{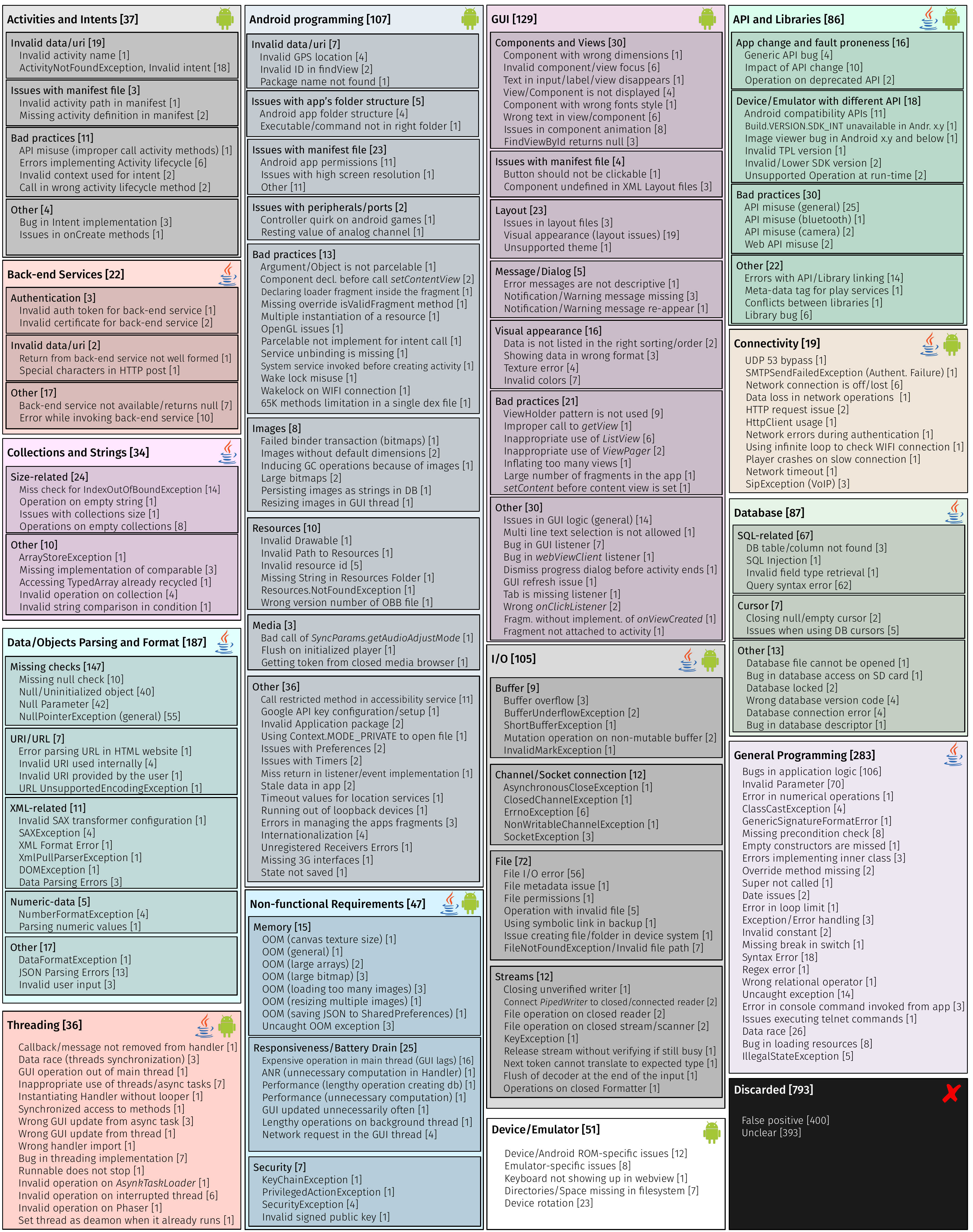}\vspace{-0.2cm}
\caption{The defined taxonomy of Android bugs.}
\label{fig:taxonomy}
\end{figure*}

\vspace{-0.3cm}
\subsection{The Defined Taxonomy}
\figref{fig:taxonomy} depicts the taxonomy that we obtained through the manual coding. The black rectangle in the bottom-right part of \figref{fig:taxonomy} reports the number of documents tagged as \emph{false positive} or as \emph{unclear}. The other rectangles---marked with the Android and/or with the Java logo---represent the 14 high-level categories that we identified. Categories marked with the Android logo (\eg Activities and Intents) group together Android-specific bugs while those marked with the Java logo (\eg Collections and Strings) group bugs that could affect any Java application. Both symbols together indicate categories featuring both Android-specific and Java-related bugs (see \eg I/O). The number reported in square brackets indicates the bug instances (from the manually classified sample) belonging to each category. Inner rectangles, when present, represent sub-categories,  \eg \emph{Responsiveness/Battery Drain} in \emph{Non-functional Requirements}.  Finally, the most fine-grained levels, represented as lighter text, describe the specific type of faults as labeled using our manually-defined tags, \eg the \emph{Invalid resource ID} tag under the sub-category {\em Resources}, in turn part of the {\em Android programming} category. The analysis of \figref{fig:taxonomy} allows to note that: 
\begin{enumerate}[leftmargin=*]
\item \emph{We were able to classify the faults reported in 1,230 documents (\eg bug reports, commits, \etc)}. This number is obtained by subtracting from the \numDocs tagged documents the 400 tagged as \emph{false positives} and the 393 tagged as \emph{unclear}. 

\item \emph{Of these 1,230, 26\% (324) are grouped in categories only reporting Android-related bugs.} This means that more than one fourth of the bugs present in Android apps are specific of this architecture, and not shared with other types of Java systems. Also, this percentage clearly represents an underestimation. Indeed, Android-specific bugs are also present in the previously mentioned ``mixed'' categories (\eg in \emph{Non-functional requirements} 25 out of the 26 instances present in the \emph{Responsiveness/Battery Drain} subcategory are Android-specific---all but \emph{Performance (unnecessary computation)}). From a more detailed count, after including also the Android-specific bugs in the ``mixed" categories,  we estimated that 35\% (430) of the identified bugs are Android-specific.

\item \emph{As expected, several bugs are related to simple Java programming}. This holds for 800 of the identified bugs (65\%). 
\end{enumerate}

\textbf{Take-away.} Over one third (35\%) of the bugs we identified with manual inspection are Android-specific. This highlights the importance of having testing instruments, such as mutation operators, tailored for such a specific type of software. At the same time, 65\% of the bugs that are typical of any Java application confirm the importance of also considering standard testing tools developed for Java, including mutation operators, when performing verification and validation activities of Android apps.


\section{Mutation Operators for Android}
\label{sec:operators}

Given the taxonomy of faults in Android apps and the set of available operators widely used for Java applications, a catalog of Android-specific mutation operators should (i) complement the classic Java operators, (ii) be representative of the faults exhibited by Android apps, (iii) reduce the rate of still-born and trivial mutants, and (iv) consider faults that can be simulated by modifying statements/elements in the app source code and resources (\eg the {\tt strings.xml} file). The last condition is based on the fact that some faults cannot be simulated by changing the source code, like in the case of device specific bugs, or bugs related to the API and third-party libraries.

Following the aforementioned conditions, we defined a set of \numMuts operators, trying to cover as many fault categories as possible (10 out of the 14 categories in \figref{fig:taxonomy}), and complementing the available Java mutation operators. The reasons for not including operators from the other four categories are: 
\begin{enumerate}[leftmargin=*]
\item API/Libraries: bugs in this category are related to API/Library issues and API misuses.  The former will require applying operators to the APIs; the latter requires a deeper analysis of the specific API usage patterns inducing the bugs;
\item Collections/Strings: most of the bugs in this category can be induced with classic Java mutation operators;
\item Device/Emulator: because this type of bug is Device/Emulator specific, their implementation is out of the scope of source code mutations;
\item Multi-threading: the detection of the places for applying the corresponding mutations is not trivial; therefore, this category will be considered in future work.
\end{enumerate}

\begin{table*}[t]

\caption{\small Proposed mutation operators. The table lists the operator names, detection strategy (\underline{AST} or \underline{TEXT}ual), the fault category (\underline{A}ctivity\underline{/I}ntents, \underline{A}ndroid \underline{P}rogramming, \underline{B}ack-\underline{E}nd \underline{S}ervices, \underline{C}onnectivity, \underline{D}ata, \underline{D}ata\underline{B}ase, \underline{G}eneral \underline{P}rogramming, \underline{GUI}, \underline{I/O}, \underline{N}on-\underline{F}unctional \underline{R}equirements), and a brief operator description. The operators indicated with * are not implemented in \mplus yet.}
\vspace{-0.4cm}
\begin{center}
\footnotesize
\begin{tabular}{p{4cm}p{0.7cm}p{0.7cm}p{10.5cm}}
\hline
\textbf{Mutation Operator}&\textbf{Det.}&\textbf{Cat.}&\textbf{Description} \\  \hline
ActivityNotDefined&Text&A/I&Delete an  activity $<$android:name=``Activity''/$>$ entry in the Manifest file\\ 
DifferentActivityIntentDefinition&AST&A/I& Replace the Activity.class argument in an Intent instantiation\\ 
InvalidActivityName&Text&A/I& Randomly insert typos in the path of an activity defined in the Manifest file\\ 
InvalidKeyIntentPutExtra&AST&A/I&Randomly generate a different key in an Intent.putExtra(key, value) call\\ 
InvalidLabel&Text&A/I&Replace the attribute ``android:label'' in the Manifest file with a random string\\ 
NullIntent&AST&A/I&Replace an Intent instantiation with null\\ 
NullValueIntentPutExtra&AST&A/I&Replace the value argument in an Intent.putExtra(key, value) call with new Parcelable[0]\\
WrongMainActivity&Text&A/I&Randomly replace the main activity definition with a different activity\\ 
\hline
MissingPermissionManifest&Text&AP&Select and remove an $<$uses-permission /$>$ entry in the Manifest file\\ 
NotParcelable&AST&AP&Select a parcelable class, remove``implements Parcelable'' and the @override annotations\\ 
NullGPSLocation&AST&AP&Inject a Null GPS location in the location services\\ 
SDKVersion&Text&AP&Randomly mutate the integer values in the SdkVersion-related attributes\\ 
WrongStringResource&Text&AP&Select a $<$string /$>$ entry in /res/values/strings.xml file and mutate the string value\\
\hline
NullBackEndServiceReturn&AST&BES&Assign null to a response variable from a back-end service\\ 
\hline 
BluetoothAdapterAlwaysEnabled&AST&C&Replace a BluetoothAdapter.isEnabled() call with``tru''\\ 
NullBluetoothAdapter&AST&C&Replace a BluetoothAdapter instance with null\\ 
\hline
InvalidURI&AST&D&If URIs are used internally, randomly mutate the URIs\\ 
\hline
ClosingNullCursor&AST&DB&Assign a cursor to null before it is closed\\ 
InvalidIndexQueryParameter&AST&DB&Randomly modify indexes/order of query parameters\\ 
InvalidSQLQuery&AST&DB&Randomly mutate a SQL query\\ 
\hline
InvalidDate&AST&GP&Set a random Date to a date object\\ 
InvalidMethodCallArgument*&AST&GP&Randomly mutate a method call argument of a basic type\\ 
NotSerializable&AST&GP&Select a serializable class, remove ``implements Serializable''\\ 
NullMethodCallArgument*&AST&GP&Randomly set null to a method call argument\\ 
\hline
BuggyGUIListener&AST&GUI&Delete action implemented in a GUI listener\\ 
FindViewByIdReturnsNull&AST&GUI&Assign a variable (returned by Activity.findViewById) to null\\ 
InvalidColor&Text&GUI&Randomly change colors in layout files\\ 
InvalidIDFindView&AST&GUI&Replace the id argument in an Activitity.findViewById call\\ 
InvalidViewFocus*&AST&GU&IRandomly focus a GUI component\\ 
ViewComponentNotVisible&AST&GUI&Set visible attribute (from a View) to false \\ 
\hline
InvalidFilePath&AST&I/O&Randomly mutate paths to files\\ 
NullInputStream&AST&I/O&Assign an input stream (\eg reader) to null before it is closed\\ 
NullOutputStream&AST&I/O&Assign an output stream (\eg writer) to null before it is closed\\ 
\hline
LengthyBackEndService&AST&NFR&Inject large delay right-after a call to a back-end service\\ 
LengthyGUICreation&AST&NFR&Insert a long delay (\ie Thread.sleep(..)) in the GUI creation thread\\ 
LengthyGUIListener&AST&NFR&Insert a long delay (\ie Thread.sleep(..)) in the GUI listener thread\\ 
LongConnectionTimeOut&AST&NFR&Increase the time-out of connections to back-end services\\ 
OOMLargeImage&AST&NFR&Increase the size of bitmaps by explicitly setting large dimensions\\  \hline
\end{tabular}
\end{center}
\label{tab:operators}
\vspace{-0.5cm}
\end{table*}

The list of defined mutation operators is provided in \tabref{tab:operators} and these operators were implemented in a tool named \mplus. In the context of this paper, we define a Potential Failure Profile (PFP) that sipulates locations of the analyzed apps---which can be source code statements, XML tags or locations in other resource files---that can be the source of a potential fault, given the faults catalog from \secref{sec:taxonomy}. Consequently, the PFP lists the locations where a mutation operator can be applied. 

In order the extract the PFP, \mplus statically analyzes the targeted mobile app, looking for locations where the operators from \tabref{tab:operators} can be implemented. The locations are detected automatically by parsing XML files or through AST-based analysis for detecting the location of API calls.  Given an automatically derived PFP for an app, and the catalog of Android-specific operators, \mplus generates a mutant for each location in the PFP. Mutants are initially generated as clones (at source code-level) of the original app, and then the clones are automatically compiled/built into individual Android Packages (APKs). Note that each location in the PFP is related to a mutation operator. Therefore, given a location entry in the PFP, \mplus automatically detects the corresponding mutation operator and applies the mutation in the source code. Details of the detection rules and code transformations applied with each operator are provided in our replication package \cite{replication}.

It is worth noting that from our catalog of Android-specific operators only two operators  (DifferentActivityIntentDefinition and MissingPermissionManifest) overlap with the eight operators proposed by Deng \etal, \cite{Deng:ICSTW15}. Future work will be devoted to cover a larger number of fault categories and define/implement a larger number of operators.

\section{Applying Mutation Testing Operators to Android Apps}
\label{sec:tools}


The \emph{goal} of this study is to: (i) understand and compare the \textit{\textbf{applicability}} of \mplus and other currently available mutation testing tools to Android apps; (ii) to understand the \textit{\textbf{underlying reasons}} for mutants---generated by these tools---that cannot be considered useful for the mutant analysis purposes, \ie mutants that do not compile or cannot be launched. This study is conducted from the \emph{perspective} of researchers interested in improving current tools and approaches for mutation testing in the context of mobile apps. 
The study addresses the following research questions:
	
\begin{itemize}[leftmargin=*]
\item{\textbf{RQ$_1$:}  \emph{Are the mutation operators (available for Java and Android apps) representative of real bugs in Android apps?}}
\item{\textbf{RQ$_2$:}  \emph{What is the rate of stillborn mutants (\eg those leading to failed compilations)  and trivial mutants (\eg those leading to crashes on app launch) produced by the studied tools when used with Android apps?}}
\item{\textbf{RQ$_3$:}  \emph{What are the major causes for stillborn and trivial mutants produced by the mutation testing tools when applied to Android apps?}}
\end{itemize}

\vspace{-0.1cm}
To answer \textbf{RQ$_1$}, we measured the applicability of operators from seven mutation testing tools (Major \cite{Just:ISSTA14}, PIT \cite{PIT}, $\mu$Java \cite{Ma:2005}, Javalanche \cite{Schuler:FSE09}, muDroid \cite{muDroid}, Deng \etal \cite{Deng:ICSTW15}, and \mplus) in terms of their ability of representing real Android apps' faults documented in a sample of software artifacts not used to build the taxonomy presented in \secref{sec:taxonomy}. To answer \textbf{RQ$_2$}, we used a representative subset of the aforementioned tools to generate mutants for 55 open source Android apps, quantitatively and qualitatively examining the stillborn and trivial mutants generated by each tool.  
Finally, to answer \textbf{RQ$_3$}, we manually analyzed the mutants and their crash outputs to qualitatively determine the reasons for trivial and stillborn mutants generated by each tool.

\subsection{Study Context and Data Collection}
\label{subsec:context}
To answer \textbf{RQ$_1$}, we analyzed the complete list of \alltoolsopers mutation operators from the seven considered tools to investigate their ability to ``\emph{cover}'' bugs described in 726 artifacts\footnote{With ``cover'' we mean the ability to generate a mutant simulating the presence of a give type of bug.} (103 exceptions hierarchy and API methods throwing exceptions, 245 bug-fixing commits from GitHub, 176 closed issues from GitHub, and 202 questions from SO). Such 726 documents were randomly selected from the dataset built for the taxonomy definition (see \secref{sub:taxdesing}) by excluding the ones already tagged and used in the taxonomy. The documents were manually analyzed by the eight authors using the same exact procedure previously described for the taxonomy building (\ie two evaluators per document having the goal of tagging the type of bug described in the document; conflicts solved by using a majority-rule schema; tagging process supported by a Web app---details in \secref{sub:taxdesing}). We targeted the tagging of $\sim$150 documents per evaluator (600 overall documents considering eight evaluators and two evaluations per document). However, some of the authors tagged more documents, leading to the considered 726 documents. Note that we did not constrain the tagging of the bug type to the ones already present in our taxonomy (\figref{fig:taxonomy}): The evaluations were free to include new types of previously unseen bugs. 

We answer RQ$_1$ by reporting (i) the new bug types we identified in the tagging of the additional 726 documents (\ie the ones not present in our original taxonomy), (ii) the \emph{coverage} level ensured by each of the seven mutation tools, measured as the percentage of bug types and bug instances identified in the 726 documents covered by its operators. We also analyze the complementarity of \mplus with respect to the existing tools.

Concerning \textbf{RQ$_2$}  and \textbf{RQ$_3$}, we compare \mplus with two popular open source mutation testing tools (Major and PIT), which are available and can be tailored for Android apps, and with one context-specific mutation testing tool for Android called muDroid \cite{Deng:ICSTW15}.  We chose these tools because of their diversity (in terms of functionality and mutation operators), their compatibility with Java, and their representativeness of tools working at different representation levels: source code, Java bytecode, and smali bytecode (\ie Android-specific bytecode representation).  

To compare the applicability of each mutation tool, we need a set of Android apps that meet certain constraints: (i) the source code of the apps must be available, (ii), the apps should be representative of different categories, and (iii) the apps should be compilable (\eg including proper versions of the external libraries they depend upon).  For these reasons, we use the Androtest suite of apps \cite{Shauvik:2015}, which includes 68 Android apps from 18 Google Play categories. These apps have been previously used to study the design and implementation of automated testing tools for Android and met the three above listed constraints. The mutation testing tools exhibited issues in 13 of the considered 68 apps, \ie the 13 apps did not compile after injecting the faults. Thus, in the end, we considered 55 subject apps in our study. The list of considered apps as well as their source code is available in our replication package~\cite{replication}.

Note that while Major and PIT are compatible with Java applications, they cannot be directly applied to Android apps. Thus, we wrote specific wrapper programs to perform the mutation, the assembly of files, and the compilation of the mutated apps into runnable Android application packages (\ie \textit{APKs}).  While the procedure used to generate and compile mutants varies for each tool, the following general workflow was used in our study: (i) generate mutants by operating on the original source/byte/smali code using all possible mutation operators; (ii) compile or assemble the APKs either using the \texttt{ant}, \texttt{dex2jar}, or \texttt{baksmali} tools; (iii) run all of the apps in a parallel-testing architecture that utilizes Android Virtual Devices (AVDs); (iv) collect data about the number of apps that crash on launch and the corresponding exceptions of these crashes which will be utilized for a manual qualitative analysis.  We refer readers to our replication package for the complete technical methodology used for each mutation tool \cite{replication}.  

To quantitatively assess the applicability and effectiveness of the considered mutation tools to Android apps, we used three metrics: \textbf{Total Number of Generated Mutants (TNGM)}, \textbf{Stillborn Mutants (SM)}, and  \textbf{Trivial Mutants (TM)}.  In this paper, we consider \textit{stillborn mutants} as those that are syntactically incorrect to the point that the APK file cannot be compiled/assembled, and \textit{trivial mutants} as those that are killed arbitrarily by nearly any test case. If a mutant crashes upon launch, we consider it as a trivial mutant. Another metric one might consider to evaluate the effectiveness of a mutation testing tool is the number of equivalent and redundant mutants the tool produces. However, in past work, the identification of equivalent mutants has been proven to be an undecidable problem \cite{equiv-mutants-1}, and both equivalent and redundant mutants require the existence of test suites (not available for the Androtest apps). Therefore, this aspect is not studied in our work.  

After generating the mutants' \texttt{APKs} using each tool, we needed a viable methodology for launching all these mutants in a reasonable amount of time to determine the number of \textit{trivial mutants}. To accomplish this, we relied on a parallel Android execution architecture that we call the Execution Engine (EE).  EE utilizes concurrently running instances of Android Virtual Devices based on the \texttt{android-x86} project~\cite{androidx86}. Specifically, we configured 20 AVDs with the \texttt{android-x86} v4.4.2 image, a screen resolution of 1900x1200, and 1GB of RAM to resemble the hardware configuration of a Google Nexus 7 device.  We then concurrently instantiated these AVDs and launched each mutant, identifying app crashes. 

\subsection{Results}


\begin{figure}[t]
\includegraphics[width=0.7\linewidth]{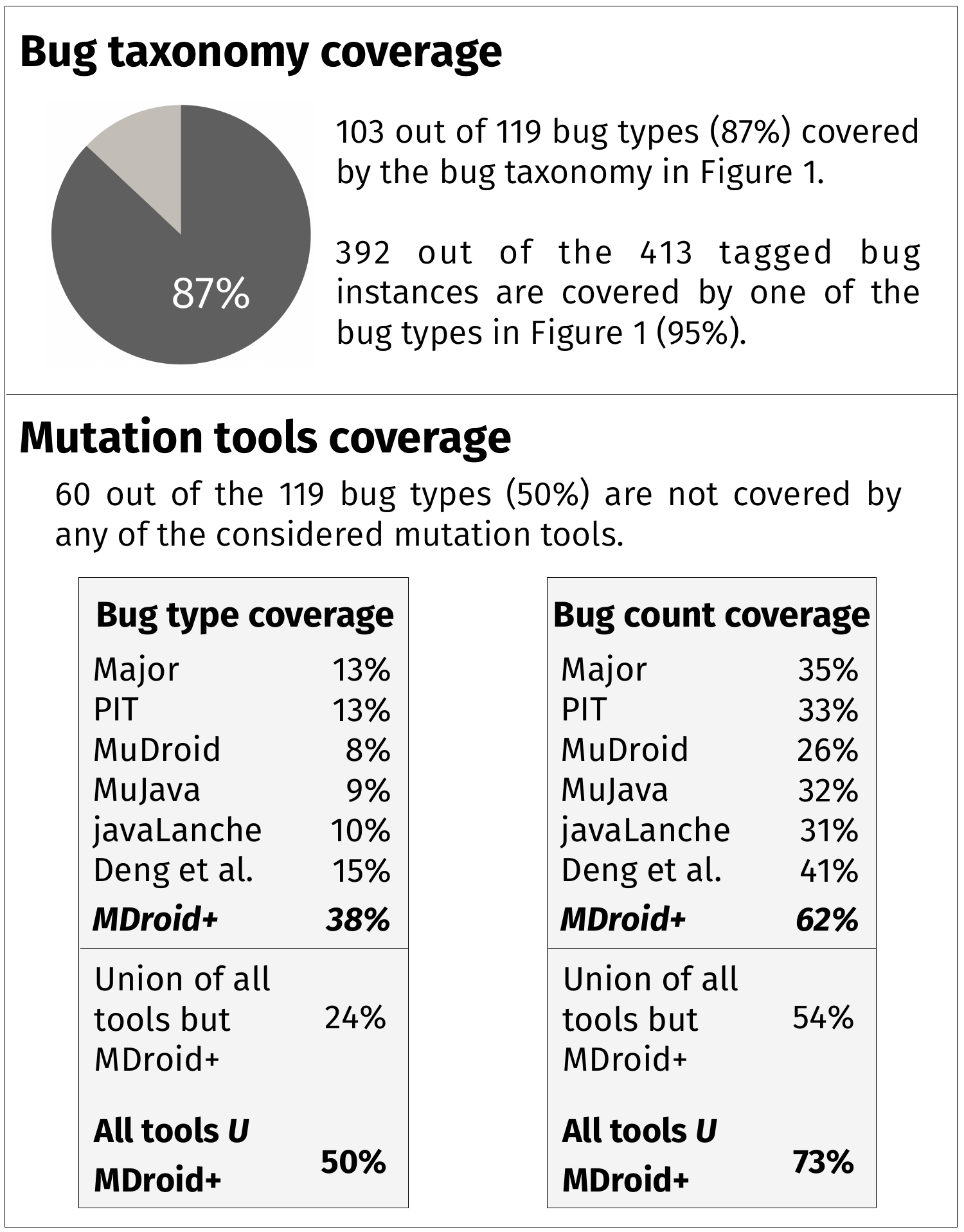}\vspace{-0.2cm}
\caption{Mutation tools and coverage of analyzed bugs.}
\label{fig:tag-second-phase}
\end{figure}

\textbf{RQ$_1$:} \figref{fig:tag-second-phase} reports (i) the percentage of bug types identified during our manual tagging that are covered by the taxonomy of bugs we previously presented in \figref{fig:taxonomy} (top part of \figref{fig:tag-second-phase}), and (ii) the coverage in terms of bug types as well as of instances of tagged bugs ensured by each of the considered mutation tools (bottom part). The data shown in \figref{fig:tag-second-phase} refers to the 413 bug instances for which we were able to define the exact reason behind the bug (this excludes the 114 entities tagged as \emph{unclear} and the 199 identified as \emph{false positives}).

87\% of the bug types are covered in our taxonomy. In particular, we identified 16 new categories of bugs that we did not encounter before in the definition of our taxonomy (\secref{sec:taxonomy}). Examples of these categories (full list in our replication package) are: \emph{Issues with audio codecs}, \emph{Improper implementation of sensors as Activities}, and \emph{Improper usage of the static modifier}. Note that these categories just represent a minority of the bugs we analyzed, accounting all together for a total of 21 bugs (5\% of the 413 bugs considered). Thus, our bug taxonomy covers 95\% of the bug instances we found, indicating a very good coverage. 

Moving to the bottom part of \figref{fig:tag-second-phase}, our first important finding highlights the limitations of the experimented mutation tools (including \mplus) in potentially unveiling the bugs subject of our study. Indeed, for 60 out of the 119 bug types (50\%), none of the considered tools is able to generate mutants simulating the bug. This stresses the need for new and more powerful mutation tools tailored for mobile platforms. For instance, no tool is currently able to generate mutants covering the \emph{Bug in webViewClient listener} and the \emph{Components with wrong dimensions} bug types. 

When comparing the seven mutation tools considered in our study, \mplus clearly stands out as the tool ensuring the highest coverage both in terms of bug types and bug instances. In particular, mutators generated by \mplus have the potential to unveil 38\% of the bug types and 62\% of the bug instances. In comparison, the best competitive tool (\ie the catalog of mutants proposed by Deng \etal \cite{Deng:ICSTW15}) covers 15\% of the bug types (61\% less as compared to \mplus) and 41\% of the bug instances (34\% less as compared to \mplus). Also, we observe that \mplus covers bug categories (and, as a consequence, bug instances) missed by all competitive tools. Indeed, while the union of the six competitive tools covers 24\% of the bug types (54\% of the bug instances), adding the mutation operators included in \mplus increases the percentage of covered bug types to 50\% (73\% of the bug instances). Examples of categories covered by \mplus and not by the competitive tools are: \emph{Android app permissions}, thanks to the MissingPermissionManifest operator, and the \emph{FindViewById returns null}, thanks to the FindViewByIdReturnsNull operator.

Finally, we statistically compared the proportion of bug types and the number of bug instances covered by \mplus, by all other techniques, and by their combination, using Fisher's exact test and Odds Ratio (OR) \cite{Sheskin2000}. The results indicate that:
\begin{enumerate}[leftmargin=*]
\item The odds of covering bug types using \mplus are 1.56 times greater than other techniques, although the difference is not statistically significant ($p$-value=0.11). Similarly, the odds of discovering faults with \mplus are 1.15 times greater than other techniques, but the difference is not significant ($p$-value=0.25);
\item The odds of covering bug types using \mplus combined with other techniques are 2.0 times greater than the other techniques alone, with a statistically significant difference ($p$-value=0.008). Similarly, the odds of discovering bugs using the combination of \mplus and other techniques are 1.35 times greater than other techniques alone, with a  significant difference ($p$-value=0.008).
\end{enumerate}


\begin{figure}[t]
	\begin{subfigure}[b]{0.225\textwidth}
		\subcaption{\%stillborn mutants}
		\vspace{-0.4cm}
		\includegraphics[width=\textwidth]{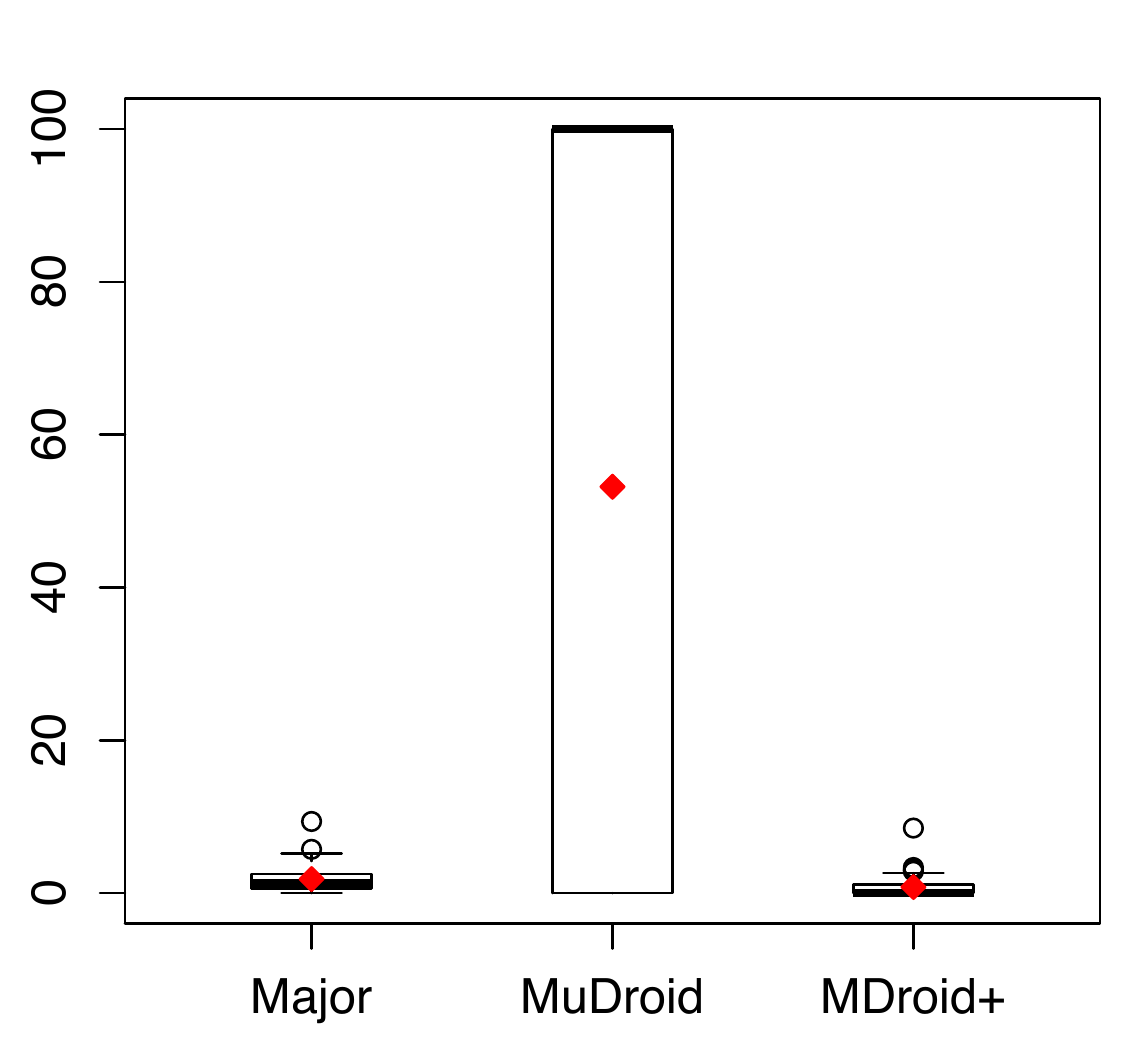}
		\label{fig:stillborn-mutants}
	\end{subfigure}
	\begin{subfigure}[b]{0.225\textwidth}
		\subcaption{\%trivial mutants}
		\vspace{-0.4cm}
		\includegraphics[width=\textwidth]{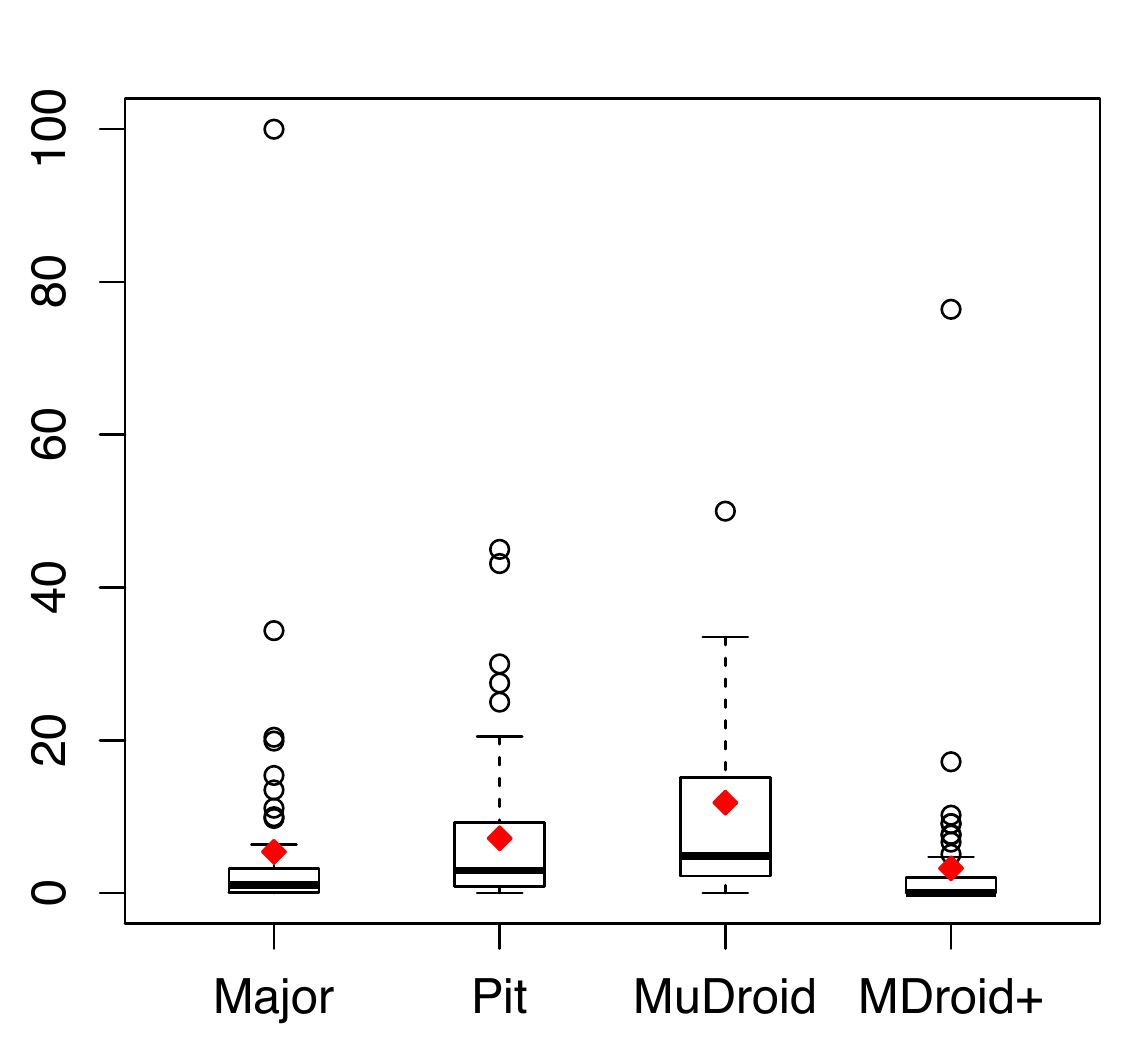}
		\label{fig:trivial-mutants}
	\end{subfigure}
	\begin{subfigure}[b]{0.46\textwidth}
		\subcaption{\# of mutants per app }
		\label{fig:gen-mutants}
		\vspace{-0.4cm}
		\includegraphics[width=\textwidth]{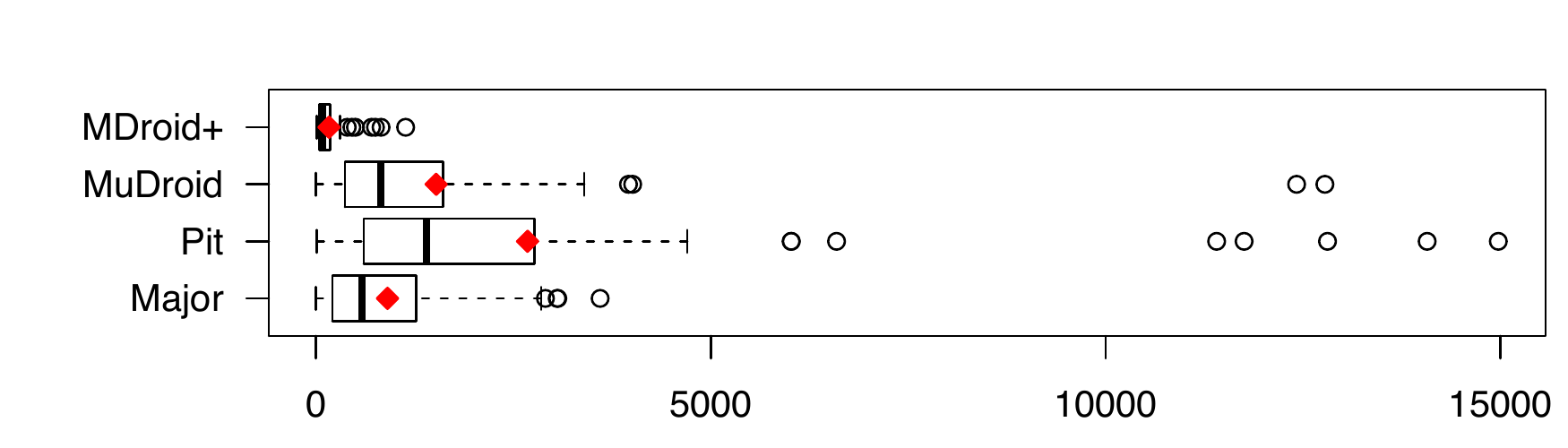}
	\end{subfigure}
	\vspace{-0.3cm}
	\caption{Stillborn and trivial mutants generated per app.}\label{fig:results}

\end{figure}

\textbf{RQ$_2$:} Figure \ref{fig:results} depicts the achieved results as percentage of (a) Stillborn Mutants (SM), and (b) Trivial Mutants (TM) generated by each tool on each app. 
On average, 167, 904, 2.6k+, and 1.5k+ mutants were generated by \mplus, Major, PIT, and muDroid, respectively for each app. The larger number of mutants generated by PIT is due in part to the larger number of mutation operators available for the tool. The average percentage of \textbf{stillborn mutants} (SM) generated by \mplus, Major and muDroid over all the apps is 0.56\%, 1.8\%, and 53.9\%, respectively, while no SM are generated by PIT (Figure~\ref{fig:stillborn-mutants}). \mplus produces significantly less SM than Major (Wilcoxon paired signed rank test  $p$-value$<0.001$ -- adjusted with Holm's correction \cite{holm}, Cliff's $d$=0.59 - large) and than muDroid (adjusted $p$-value$<0.001$, Cliff's $d$=0.35 - medium).

These differences across the tools are mainly due to the compilation/assembly process they adopt during the mutation process. PIT works at Java bytecode level and thus can avoid the SM problem, at the risk of creating a larger number of TM. However, PIT is the tool that required the highest effort to build a wrapper to make it compatible with Android apps. Major works at the source code level and compiles the app in a ``traditional" manner. Thus, it is prone to SM and requires an overhead in terms of memory and CPU resources needed for generating the mutants. Finally, muDroid operates on \texttt{APKs} and \texttt{smali} code, reducing the computational cost of mutant generation, but significantly increasing the chances of SM. 

All four tools generated \textbf{trivial mutants} (TM) (\ie mutants that crashed simply upon launching the app). These instances place an unnecessary burden on the developer, particularly in the context of mobile apps, as they must be discarded from analysis.  
The mean of the distribution of the percentage of TM over all apps for \mplus, Major, PIT and muDroid is 2.42\%, 5.4\%, 7.2\%, and 11.8\%, respectively (Figure~\ref{fig:trivial-mutants}). \mplus generates significantly less TM than muDroid (Wilcoxon paired signed rank test adjusted $p$-value=0.04, Cliff's $d$=0.61 - large) and than PIT (adjusted $p$-value=0.004, Cliff's $d$=0.49 - large), while there is no statistically significant difference with Major (adjusted $p$-value=0.11).

While these percentages may appear small, the raw values show that the TM can comprise a large set of instances for tools that can generate thousands of mutants per app. For example, for the Translate app, 518 out of the 1,877 mutants generated by PIT were TM. For the same app, muDroid creates 348 TM out of the 1,038 it generates.  For the Blokish app, 340 out of the 3,479 mutants generated by Major were TM. Conversely, while \mplus may generate a smaller number of mutants per app, this also leads to a smaller number of TM, only 213 in total across all apps. This is due to the fact that \mplus generates a much smaller set of  mutants that are specifically targeted towards emulating \textit{real} faults identified in our empirically derived taxonomy, and are applied on specific locations detected by the PFP.

\textbf{RQ$_3$:}  In terms of mutation operators causing the highest number of stillborn and TM we found that for Major, the Literal Value Replacement (LVR) operator had the highest number of TM, whereas the Relational Operator Replacement (ROR) had the highest number of SM. It may seem surprising that ROR generated many SM, however, we discovered that the reason was due to improper modifications of loop conditions.  For instance, in the A2dp.Vol app one mutant changed this loop: \texttt{for (int i = 0; i $<$ cols; i++)} and replaced the condition ``\texttt{$i$ $<$ $cols$}" with ``\texttt{false}", causing the compiler to throw an unreachable code error.  For PIT, the Member Variable Mutator (MVM) is the one causing most of the TM; for muDroid, the Unary Operator Insertion (UOI) operator has the highest number of SM (although all the operators have relatively high failure rates), and the Relational Value Replacement (RVR) has the highest number of TM. For \mplus, the WrongStringResource operator had that highest number of SM, whereas the FindViewByIdReturnsNull operator had the highest number of TM.




\begin{table}[]
\centering
\caption{Number of Generated, Stillborn, and Trivial Mutants created by MDroid+ operators.}
\label{tab:mplus-stats}
\footnotesize
\vspace{-0.4cm}
\begin{tabular}{|l|l|l|l|}
\hline
\textbf{Mutation Operators}                & \textbf{GM} & \textbf{SM} & \textbf{TM} \\ \hline
WrongStringResource               & 3394 & 0  & 14 \\ \hline
NullIntent                        & 559  & 3  & 41 \\ \hline
InvalidKeyIntentPutExtra          & 459  & 3  & 11 \\ \hline
NullValueIntentPutExtra           & 459  & 0  & 14 \\ \hline
InvalidIDFindView                 & 456  & 4  & 30 \\ \hline
FindViewByIdReturnsNull           & 413  & 0  & 40 \\ \hline
ActivityNotDefined                & 384  & 1  & 8  \\ \hline
InvalidActivityName               & 382  & 0  & 10 \\ \hline
DifferentActivityIntentDefinition & 358  & 2  & 8  \\ \hline
ViewComponentNotVisible           & 347  & 5  & 7  \\ \hline
MissingPermissionManifest         & 229  & 0  & 8  \\ \hline
InvalidFilePath                   & 220  & 0  & 1  \\ \hline
InvalidLabel                      & 214  & 0  & 3  \\ \hline
ClosingNullCursor                 & 179  & 13 & 5  \\ \hline
LengthyGUICreation                & 129  & 0  & 1  \\ \hline
BuggyGUIListener                  & 122  & 0  & 2  \\ \hline
LengthyGUIListener                & 122  & 0  & 0  \\ \hline
SDKVersion                        & 66   & 0  & 2  \\ \hline
NullInputStream                   & 61   & 0  & 4  \\ \hline
WrongMainActivity                 & 56   & 0  & 0  \\ \hline
InvalidColor                      & 52   & 0  & 0  \\ \hline
NullOuptutStream                  & 45   & 0  & 2  \\ \hline
InvalidDate                       & 40   & 0  & 0  \\ \hline
InvalidSQLQuery                   & 33   & 0  & 2  \\ \hline
NotSerializable                   & 15   & 7  & 0  \\ \hline
NullBluetoothAdapter              & 9    & 0  & 0  \\ \hline
LengthyBackEndService             & 8    & 0  & 0  \\ \hline
NullBackEndServiceReturn          & 8    & 1  & 0  \\ \hline
NotParcelable                     & 7    & 6  & 0  \\ \hline
InvalidIndexQueryParameter        & 7    & 1  & 0  \\ \hline
OOMLargeImage                     & 7    & 4  & 0  \\ \hline
BluetoothAdapterAlwaysEnabled     & 4    & 0  & 0  \\ \hline
InvalidURI                        & 2    & 0  & 0  \\ \hline
NullGPSLocation                   & 1    & 0  & 0  \\ \hline
LongConnectionTimeOut             & 0    & 0  & 0  \\ \hline
\textbf{Total} & 8847 & 50 & 213\\ \hline
\end{tabular}

\end{table}

To qualitatively investigate the causes behind the crashes, three authors manually analyzed a randomly selected sample of 15 crashed mutants per tool. In this analysis, the authors relied on information about the mutation (\ie applied mutation operator and location), and the generated stack trace. 

\textbf{Major.} The reasons behind the crashing mutants generated by Major mainly fall in two categories. First, mutants generated with the LVR operator that changes the value of a literal causing an app to crash. This was the case for the \emph{wikipedia} app when changing the ``1'' in the invocation {\tt set\-Cache\-Mode\-(params.getString(1))} to ``0''. This passed a wrong asset URL to the method {\tt set\-Cache\-Mode}, thus crashing the app. Second, the Statement Deletion (STD) operator was responsible for app crashes especially when it deleted needed methods' invocations. A representative example is the deletion of invocations to methods of the superclass when overriding methods, \eg when removing the {\tt super.onDestroy()} invocation from the {\tt onDestroy()} method of an {\tt Activity}. This results in throwing of an {\tt android.\-util.\-Super\-Not\-Called\-Exception}. Other STD mutations causing crashes involved deleting a statement initializing the main {\tt Activity} leading to a {\tt Null\-Pointer\-Exception}.

\textbf{muDroid.} This tool is the one exhibiting the highest percentage of stillborn and TM. The most interesting finding of our qualitative analysis is that 75\% of the crashing mutants lead to the throwing of a {\tt java.\-lang.\-VerifyError}. A {\tt VerifyError} occurs when Android tries to load a class that, while being syntactically correct, refers to resources that are not available (\eg wrong class paths). In the remaining 25\% of the cases, several of the crashes were due to the Inline Constant Replacement (ICR) operator. An example is the crash observed in the {\tt photostream} app where the ``100'' value has been replaced with ``101'' in {\tt bitmap.\-compress\-(Bitmap.\-Compress\-Format.\-PNG, \-100, \-out)}. Since ``100'' represents the quality of the compression, its value must be bounded between 0 and 100. 

\textbf{PIT.} In this tool, several of the manually analyzed crashes were due to (i) the RVR operator changing the return value of a method to null, causing a {\tt Null\-Pointer\-Exception}, and (ii) removed method invocations causing issues similar to the ones described for Major.

\textbf{MDroid+.} \tabref{tab:mplus-stats} lists the mutants generated by \mplus across all the systems (information for the other tools is provided with our replication package). 
The overall rate of SM is very low in \mplus, and most failed compilations pertain to edge cases that would require a more robust static analysis approach to resolve.  For example, the ClosingNullCursor operator has the highest total number of SM (across all the apps) with 13, and some edge cases that trigger compilation errors involve cursors that have been declared \texttt{Final}, thus causing the reassignment to trigger the compilation error.  The small number of other SM are generally other edge cases, and current limitations of \mplus can be found in our replication package with detailed documentation. 

The three operators generating the highest number of TM are NullIntent(41), FindViewByIdReturnsNull(40), and InvalidIDFindView(30).  The main reason for the NullIntent TM are intents invoked by the Main Activity of an app (\ie the activity loaded when the app starts). Intents are one of the fundamental components of Android apps and function as asynchronous messengers that activate Activities, Broadcast Receivers and services.  One example of a trivial mutant is for the A2dp.Vol app, in which a bluetooth service, inteneded to start up when the app is launched, causes a NullPointerException when opened due to NullIntent operator.  To avoid cases like this, more sophisticated static analysis could be performed to prevent mutations from affecting Intents in an app's MainActivity.  The story is similar for the FindViewViewByIdReturnsNull and InvalidIDFindView operators:  TM will occur when views in the MainActivity of the app are set to null or reference invalid Ids, causing a crash on startup. Future improvements to the tool could avoid mutants to be seeded in components related to the MainActivity. Also, it would be desirable to allow developers to choose the activities in which mutations should be injected.  

\emph{\textbf{Summary of the RQs}. \mplus outperformed the other six mutation tools by achieving the highest coverage both in terms of bug types and bug instances. However, the results show that Android-specific mutation operators should be combined with classic operators to generate mutants that are representative of real faults in mobile apps (\textbf{RQ$_1$}). \mplus generated the smallest rate of both stillborn and trivial mutants illustrating its immediate applicability to Android apps.  Major and muDroid generate stillborn mutants, with the latter having a critical average rate of 58.7\% stillborn mutants per app (\textbf{RQ$_2$}). All four tools generated a relatively low rate of trivial mutants, with muDroid again being the worst with an 11.8\% average rate of trivial mutants (\textbf{RQ$_3$}). Our analysis shows that the PIT tool is most applicable to Android apps when evaluated in terms of the ratio between failed and  generated mutants. However, \mplus is both practical and based on Android-specific operations implemented according to an empirically derived fault-taxonomy of Android apps.}



\section{Threats to Validity}
\label{sec:threats}
This section discusses the threats to validity of the work related to devising the fault taxonomy, and carrying out the study reported in \secref{sec:tools}.

Threats to {\em construct validity} concern the relationship between theory and observation. The main threat is related to how we assess and compare the performance of mutation tools, \ie by covering the types, and by their capability to limit stillborn and trivial mutants. A further, even more relevant evaluation would explore the extent to which different mutant taxonomies are able to support test case prioritization. However, this requires a more complex setting which we leave for our future work.

Threats to {\em internal validity} concern factors internal to our settings that could have influenced our results. This is, in particular,  related to possible subjectiveness of mistakes in the tagging of \secref{sec:taxonomy} and for {\bf RQ$_1$}. As explained, we employed multiple taggers to mitigate such a threat.

Threats to {\em external validity} concern the generalizability of our findings. To maximize the generalizability of the fault taxonomy, we have considered six different data sources. However, it is still possible that we could have missed some fault types available in sources we did not consider, or due to our sampling methodology. Also, we are aware that in our study results of {\bf RQ$_1$} are based on the new sample of data sources, and results of {\bf RQ$_2$} on the set of 68 apps considered \cite{Shauvik:2015}.

\vspace{-0.1cm}
\section{Conclusions}
\label{sec:concl}
Although Android apps rely on the  Java language as a programming platform, they have specific elements that make the testing process different than other Java applications. In particular, the type and distribution of faults exhibited by Android apps may be very peculiar, requiring, in the context of mutation analysis, specific operators.

In this paper, we presented the first taxonomy of faults in Android apps, based on a manual analysis of \numDocs software artifacts from six different sources. The taxonomy is composed of \numCategs categories containing \numFaults types. Then, based on the taxonomy, we have defined
 a set of \numMuts Android-specific mutation operators, implemented in an infrastructure called \mplus, to automatically seed mutations in Android apps. To validate the taxonomy and \mplus, we conducted a comparative study with Java mutation tools. The study results show that \mplus operators are more representative of Android faults than other catalogs of mutation operators, including both Java and Android-specific operators previously proposed.  Also \mplus is able to outperform state-of-the-art tools in terms of stillborn and trivial mutants. 
 
The obtained results make our taxonomy and \mplus ready to be used and possibly extended by other researchers/practitioners. To this aim, \mplus and the wrappers for using Major and Pit with Android apps are available as open source projects \cite{replication}. Future work will extend \mplus by implementing more operators, and creating a framework for mutation analysiss. Also, we plan to experiment with \mplus in the context of test case prioritization.

\section*{Acknowledgments} Bavota was supported in part by the SNF project JITRA, No. 172479.

\balance
\bibliographystyle{ACM-Reference-Format}
\bibliography{main}
\end{document}